\documentclass[pss]{wiley2sp} 
\usepackage{amsmath}
\usepackage{bm}              

\begin{document}

\title{Current-induced spin polarization for a general two-dimensional electron system}

\titlerunning{Current-induced spin polarization for a general two-dimensional electron system}

\author{%
  C. M. Wang\textsuperscript{\Ast,\textsf{\bfseries 1}},
  H. T. Cui\textsuperscript{\textsf{\bfseries 1}},
  and Q. Lin\textsuperscript{\textsf{\bfseries 2}}}

\authorrunning{C. M. Wang et al.}

\mail{e-mail
  \textsf{cmwangsjtu@gmail.com}}

\institute{%
  \textsuperscript{1}\,Department of Physics,
Anyang Normal University, Anyang 455000, China\\
  \textsuperscript{2}\,Shanghai University of
Engineering Science, 333 Longteng Road, Songjiang, Shanghai 201620,
China}


\abstract{%
%
%
%
\abstcol{%
  In this paper, current-induced spin polarization for two-dimensional electron gas with a general spin-orbit
interaction is investigated. For isotropic energy spectrum, the
in-plane current-induced spin polarization is found to be dependent
on the electron density for non-linear spin-orbit interaction and
increases with the increment of sheet density, in contrast to the
case for $\bm k$-linear spin-orbit coupling model. The numerical
evaluation is performed for InAs/InSb heterojunction with spin-orbit
coupling of both linear and cubic spin-orbit coupling types.
 For $\delta$-type
  short-range electron-impurity scatteri-}{ng, it is found that the current-induced spin
  polarization increases with increasing the density when cubic
  spin-orbit couplings are considered. However, for remote
  disorders, a rapid enhancement of current-induced spin polarization is
  always observed at high electron density, even in the case without cubic spin-orbit
  coupling. This result demonstrates the collision-related feature of current-induced
  spin polarization. The effects of different high order spin-orbit couplings on spin
polarization can be comparable.}}

%
%

\maketitle   

\section{introduction}
Spintronics, where the spin degree of freedom is manipulated to
control the electronic devices, has been becoming a rapid field of
condensed matter physics \cite{Zutic}. Current-induced spin
polarization (CISP) discloses the possibility of the spin
polarization generated in semiconductor directly by the electric
field. It refers to a spatially homogeneous spin polarization in two
dimensional systems due to an in-plane charge current. CISP due to
spin-dependent scattering was first reported by D'yakonov and Perel'
in 1971 \cite{dyakonov1971cis}. Later it was realized that such
transport phenomenon exists in semiconductor-based two-dimensional
electron gas (2DEG) without structure or bulk inversion symmetry
\cite{edelstein1990spc,aronov,Inoue2003}. Experimentally, CISP was
first measured by Silov {\it et al.} in two dimensional hole system
with the help of polarized photoluminescence technology
\cite{silov2004cis}. Observations of CISP in strained semiconductors
have been reported by Kato \cite{Kato2004n,PhysRevLett93176601}. And
later Sih {\it et al.} demonstrated the existence of CISP in AlGaAs
quantum well \cite{sih2005mn}.

So far, the theoretical investigations about the CISP are mainly
focused on the 2DEG with $\bm k$-linear SOC, such as Rashba SOC due
to structure inversion asymmetry
$H_R=\alpha(k_y\sigma_x-k_x\sigma_y)$
\cite{edelstein1990spc,Inoue2003,bryksin2006tef}, linear Dresselhaus
SOC due to bulk inversion asymmetry
$H_D^{(1)}=\beta(k_x\sigma_x-k_y\sigma_y)$ \cite{trushin155323}, and
the combination of linear Rashba and Dresselhaus coupling types
\cite{Chaplik744,Averkiev,bryksin2006}. Here $\bm
\sigma=(\sigma_x,\sigma_y,\sigma_z)$ represents the set of Pauli
matrices, and $\bm k=(k_x,k_y)=k(\cos\theta,\sin\theta)$ is the
two-dimensional momentum, $\alpha$ and $\beta$ are linear Rashba and
Dresselhaus SOC factors, respectively. CISP is found to be
proportional to the SOC constant and independent of the electron
density for 2DEG with $\bm k$-linear Rashba or Dresselhaus SOC when
a dc electric field is applied
\cite{edelstein1990spc,Inoue2003,trushin155323}.

However situations may become different for 2DEG with non-linear
SOC, just like the spin Hall effect
\cite{PhysRevLett831834,ShuichiMurakami09052003,sinova2004uis,Bernevig,malshukov2005shc,murakami241202,shytov2006,lin2006she},
the investigation of CISP on this system is desirable. Liu {\it et
al.} studied the CISP in hole-doped two dimensional system lacking
structure inversion symmetry, and they found that CISP is dependent
on the Fermi energy, in vivid contrast against the case in 2DEG with
$\bm k$-linear SOC \cite{liu125345}. While it was pointed by Liu
{\it et al.} that the ``spin" for hole system is actually the total
angular momentum, where the spin of hole system is not a conserved
physical quantity \cite{liu125345}. Hence, we expect the behavior of
the spin polarization of {\it conduction electron} in the presence
of non-linear SOC. We know that at high electron sheet density, the
cubic term of Dresselhaus SOC,
$H_D^{(3)}=\eta(k_xk_y^2\sigma_x-k_x^2k_y\sigma_y)$
\cite{dresselhaus1955soc}, ($\eta$ is the cubic Dresselhaus SOC
constant), has to be taken into account \cite{jusserand1992zmf}.
 Recently, by using the double-group representations, Cartoix\`{a} {\it et al.} \cite{Cartoixa2006} found
   that there is another cubic term for heterojunction due to bulk inversion
   asymmetry,
    $H_{BIA}^{(3)}=\zeta(k_y^3\sigma_y-k_x^3\sigma_x)$, ({\it i. e.} the last term of Eq. (3)
    in Ref. \cite{Cartoixa2006}). Here $\zeta=a^2\beta/6$ is the cubic SOC
constant with $a$ as the well width. Apart from the type of SOC,
most of the pioneering researches on CISP treat the
electron-impurity collision with a simple momentum-independent form,
described by a relaxation time $\tau$.
  However in realistic 2DEG, the electron density is not large
  enough to screen the charged impurities. The interaction between
  electron and disorder is long ranged.

In this paper we consider the CISP in 2DEG with a general SOC, which
can be applied to describe Rashba, linear and cubic Dresselhaus
SOCs, and many other SOCs. For the simple isotropic energy band
form, the analytical result of CISP is obtained. The numerical
evaluation for the electron system with both linear and cubic SOCs
due to bulk inversion asymmetry is also performed, considering both
short- and long-range disorders.

\section{formalism}
We consider a two-dimensional non-interacting electron system with a
general SOC, described by the following one-particle Hamiltonian:
\begin{equation}\label{ham}
\hat H=\varepsilon_0(\bm k)+\bm  b(\bm k)\cdot \bm\sigma.
\end{equation}
For simplicity we have assumed that $\varepsilon_0(\bm k)$, the
energy dispersion in the absence of SOC, is isotropic function of
momentum $\bm k$. We set $\hbar=1$ throughout this paper. As a
further simplification of SOC, we shall later specialize to the
model, where the spin orbit field $\bm b(\bm k)$ has the form
\begin{equation}\label{socf}
b_x(\bm k)+ib_y(\bm k)=\tilde{\alpha}k^{M_1}(\sin M_2\theta)^\lambda
e^{iM_3\theta}.
\end{equation}
Here the complex number $\tilde{\alpha}=\alpha_r+i\alpha_i$ is the
general coupling constant, irrespective of the momentum $\bm k$,
with $\alpha_r$ and $\alpha_i$ as the real and imaginary part,
respectively. $\lambda=0,1$ and $M_1,M_2,M_3$ are integer numbers.
It will be noted later that the index $\lambda$ determines whether
the energy spectrum is isotropic. The number $M_1$ usually is
positive. It is found that when $\lambda=0$ and $M_1=M_3$, our model
becomes the one in Ref. \cite{shytov2006}, where the model is used
to discuss the spin Hall effect. Due to time reversal
 symmetry, the
spin orbit field $\bm b(\bm k)$ satisfies $\bm b(\bm k)=-\bm b(-\bm
k)$. Therefore for the case $\lambda=0$, {\it i. e. } isotropic
energy spectrum, $M_3$ must be an odd integer number; while for the
case $\lambda=1$, {\it i. e. } anisotropic energy spectrum,
$M_2+M_3$ must be an odd integer number. Now we list some special
SOC forms: for pure Rashba SOC, $\tilde \alpha=-i\alpha$,
$\lambda=0$, $M_1=M_3=1$; while for $\bm k$-linear Dresselhaus SOC,
$\tilde \alpha=\beta$, $\lambda=0$, $M_1=1$, $M_3=-1$; and the case
$\tilde \alpha=-\frac{1}{2}i\eta$, $\lambda=1$, $M_1=3$, $M_2=2$,
$M_3=1$ corresponds the cubic Dresselhaus term.

With the help of the local unitary matrix
\begin{equation}\label{}
U_{\bm k}=\frac{1}{\sqrt{2}}\left(
                              \begin{array}{cc}
                                1 & 1 \\
                                ie^{i\chi_{\bm k}} & -ie^{i\chi_{\bm k}} \\
                              \end{array}
                            \right),
\end{equation}
where $\chi_{\bm k}$ satisfies
\begin{equation}\label{}
\tan{\chi_{\bm k}}=\frac{\alpha_i \sin M_3\theta-\alpha_r \cos
M_3\theta}{\alpha_i \cos M_3\theta+\alpha_r \sin M_3\theta},
\end{equation}
the Hamiltonian \eqref{ham} can be diagonalized into $H={\rm
diag}[\varepsilon_{ 1}(\bm k),\varepsilon_{2}(\bm k)]$ in the
helicity basis. Here
\begin{equation}\label{}
\varepsilon_{\mu}(\bm k)=\varepsilon_0(\bm
k)+(-1)^\mu\varepsilon_{M}(\bm k),
\end{equation}
with $\varepsilon_{M}(\bm k)=|\tilde{\alpha}|k^{M_1}(\sin
M_2\theta)^\lambda$ and $\mu=1,2$ as the helix band index. We note
that when $\lambda=0$, the energy dispersion $\varepsilon_{\mu}(\bm
k)$ becomes isotropic function of wave vector $\bm k$, whereas for
the case $\lambda=1$, the energy spectrum relies on the angle of
momentum.

When the system is driven by a weak dc electric field applied along
the $\hat x$ direction, $\bm E=E_0\hat x$. Following the procedure
of Ref. \cite{liu2005des}, the kinetic equation of the distribution
function $\rho(\bm k)$ can be derived, with the equilibrium
distribution function
\begin{equation}\label{eqli}
\rho^{(0)}=\left(
             \begin{array}{cc}
               n_{\rm F}[\varepsilon_{1}(\bm k)] & 0 \\
               0 &  n_{\rm F}[\varepsilon_{2}(\bm k)] \\
             \end{array}
           \right).
\end{equation}
Here $n_{\rm F}(x)$ is the Fermi distribution. The distribution
function $\rho(\bm k)$ to first order of electric field comprises
two terms. The first term is written as
\begin{equation}\label{rho1}
\rho_{12}^{(1)}=\rho_{21}^{(1)}=-\frac{e E_0}{4\varepsilon_{M}(\bm
k)}\frac{\partial \chi_{\bm k}}{\partial k_x}\Big\{n_{\rm
F}[\varepsilon_{1}(\bm k)]-n_{\rm F}[\varepsilon_{2}(\bm k)]\Big\}.
\end{equation}
And the second term $\rho^{(2)}(\bm k)$ is determined by the set of
equations with the form
\begin{align}
  eE_0\frac{\partial n_{\rm F}[\varepsilon_{\mu}(\bm k)]}{\partial k_x}&=\pi\sum_{\bm q\mu'}\vert u(\bm k-\bm q)\vert^2
  \Omega_{\mu\mu'}\nonumber\\\times&\left[\rho_{\mu\mu}^{(2)}(\bm k)-
  \rho_{\mu'\mu'}^{(2)}(\bm q)\right]\delta(\varepsilon_{\mu}(\bm k)-\varepsilon_{\mu'}(\bm q)), \label{eq1}\\
  4\varepsilon_{M}(\bm k){\rm Re}\rho_{12}^{(2)}(\bm k)&=\pi\sum_{\bm q\mu\mu'}\vert u(\bm k-\bm q)\vert^2
  \bar{\Omega}_{\mu\mu'}\nonumber\\\times&\left[\rho_{\mu\mu}^{(2)}(\bm k)-\rho_{\mu'\mu'}^{(2)}(\bm q)
  \right]\delta(\varepsilon_{\mu}(\bm k)-\varepsilon_{\mu'}(\bm q)).\label{eq2}
\end{align}
Here $\Omega_{\mu\mu'}=1+(-1)^{\mu+\mu'}\cos(\chi_{\bm k}-\chi_{\bm
q})$ and $\bar{\Omega}_{\mu\mu'}=(-1)^{\mu'}\sin(\chi_{\bm
k}-\chi_{\bm q})$. ${\rm Re}\rho_{12}^{(2)}(\bm k)$ represents the
real part of the off-diagonal distribution function
$\rho_{12}^{(2)}(\bm k)$. $u(\bm k)$ is the electron-impurity
scattering matrix. Note that the above equations of distribution
function, Eqs. \eqref{eq1} and \eqref{eq2}, have been derived in
Refs. \cite{edelstein1990spc,aronov,Averkiev}.

Finally, in helix spin basis, the single-particle operators of spin
polarization are given by $\hat S_i=U_{\bm k}^\dag\frac{1}{2}{{
\sigma_i}}U_{\bm k}$ with $i=x,y,z$. The corresponding macroscopical
quantities are obtained by taking the statistical average over them,
$S_i=\sum_{\bm k}{\rm Tr}[\rho(\bm k)\hat S_i]$, and are expressed
as
\begin{eqnarray}
  S_x &=& \frac{1}{2}\sum_{\bm k\mu}\sin\chi_{\bm k}[\rho_{22}(\bm k)-\rho_{11}(\bm k)], \label{sxm}\\
  S_y &=& \frac{1}{2}\sum_{\bm k\mu}\cos\chi_{\bm k}[\rho_{22}(\bm k)-\rho_{11}(\bm k)], \label{sym}\\
  S_z &=& \sum_{\bm k}{\rm Re}\rho_{12}(\bm k).\label{szm}
\end{eqnarray}

\section{spin polarization}
\subsection{analytical result}
For this angle-dependent SOC, the formulas to be derived will become
much less transparent, and the integrals are more difficult to solve
analytically. Therefore, we first limit ourselves to isotropic
energy spectrum, {\it i. e.} $\lambda=0$ and the parabolic case
$\varepsilon_0(\bm k)=\frac{k^2}{2m}$. The spin polarization is
examined in the presence of electron-impurity scattering with
$\delta$-potential, $|u(\bm k-\bm q)|^2=n_iu_0^2$. Here $m$ is the
effective mass of two-dimensional electrons and $n_i$ is the
impurity density.

Keeping only the lowest-order of spin-orbit interaction, the
diagonal elements of distribution function can be obtained
analytically from Eq. \eqref{eq1}. For $M_3=\pm1$, the diagonal
elements of $\rho^{(2)}(\bm k)$ take the form
\begin{eqnarray}
  \rho_{11}^{(2)}(\bm k)=& -&\frac{eE_0\tau}{m}\left[k+m|\tilde{\alpha}|(1-M_1)(2\pi N)^\frac{M_1-1}{2}\right]
  \nonumber\\&\times&\cos\theta \delta(\varepsilon_{\bm k1}-\varepsilon_F),\\
  \rho_{22}^{(2)}(\bm k)=&-&\frac{eE_0\tau}{m}\left[k-m|\tilde{\alpha}|(1-M_1)(2\pi N)^\frac{M_1-1}{2}\right]
  \nonumber\\&\times&\cos\theta \delta(\varepsilon_{\bm
  k2}-\varepsilon_F).
\end{eqnarray}
And for the case $M_3>1$ or $M_3<-1$,  they are given by
\begin{eqnarray}
  \rho_{11}^{(2)}(\bm k)&=& -eE_0\tau\left|\frac{\partial \varepsilon_1(\bm k)}{\partial k}\right|
  \cos\theta \delta(\varepsilon_{\bm k1}-\varepsilon_F),\\
  \rho_{22}^{(2)}(\bm k)&=& -eE_0\tau\left|\frac{\partial \varepsilon_2(\bm k)}{\partial k}\right|
  \cos\theta \delta(\varepsilon_{\bm
  k2}-\varepsilon_F),
\end{eqnarray}
 with $\tau=1/mn_iu_0^2$ as the
relaxation time, $\varepsilon_F$ as the Fermi energy. With the help
of Eq. \eqref{eq2}, the off-diagonal distribution function can be
obtained directly. Finally, we find the spin polarization takes the
form:
\begin{eqnarray} \label{eq:3}
 S_x &=&\left\{ \begin{aligned}
         &\frac{em\alpha_r\tau E_0}{2\pi}(2\pi N)^\frac{M_1-1}{2},&|M_3|=1\\
         &0,&|M_3|>1
                \end{aligned} \right.,
\\
 S_y &=&\left\{ \begin{aligned}
         &\frac{em\alpha_i\tau E_0}{2\pi}(2\pi N)^\frac{M_1-1}{2},&|M_3|=1\\
         &0,&|M_3|>1
                \end{aligned} \right.,
\\
S_z&=&0.
\end{eqnarray}
We come to the conclusion that the in-plane CISPs exist only when
the winding number $|M_3|=1$, nevertheless the out-of-plane
component of CISP is always zero for this 2DEG with isotropic
general SOC. As expected, for $M_3=\pm1$ and $M_1=1$, we obtain the
CISP for 2DEG with Rashba or $\bm k$-linear Dresselhaus SOC, in
agreement with previous theoretical studies
\cite{edelstein1990spc,Inoue2003,trushin155323,Chaplik744,Averkiev}.
Our results imply that, in contrast to the 2DEG with Rashba or $\bm
k$-linear Dresselhaus SOC, the in-plane CISPs for non-linear SOC
system, $M_1>1$, become dependent on the electron density and
enhance for high density semiconductor. The above analytical
calculation is valid in the weak SOC case, $\varepsilon_{M}(k_{\rm
F})\ll\varepsilon_0(k_{\rm F})$, but the relationship between
$\varepsilon_{M}(k_{\rm F})$ and $\tau^{-1}$ is arbitrary. Here
$k_{\rm F}$ is the Fermi wave vector. For strong SOC case, the
dependence of CISP on the electron density can be evaluated
numerically.

 It should be
noted that the in-plane CISP comes from the interband processes,
arising from the SOC, which can be seen from Eqs. \eqref{sxm} and
\eqref{sym}. In the absence of spin-orbit interaction,
$\rho_{11}=\rho_{22}$ leads to the vanishing CISP. Hence, although
the result, we obtained here, is for the parabolic case, the
nonparabolic contribution of $\varepsilon_0(\bm k)$ to CISP may only
change its value slightly through the Fermi energy. Further, it can
be confirmed below by the numerical calculation. In general, SOC
field will be the combination of Eq. \eqref{socf}
  with both linear and high order terms. From our analytical result, one can deduce that with increasing the high order SOC constant,
  the density-related feature of CISP will become more and more evident.

\subsection{numerical result}

 Now we perform the numerical evaluation for CISP in InAs/InSb
heterojunctions without the additional large bias voltage, where the
main SOC contribution terms arise owing to the absence of the center
of inversion in the bulk material.

Further, to take account of the nonparabolicity of the energy band
of InAs, we use the isotropic Kane band model:
\begin{equation}
\varepsilon_0(k)=\frac{1}{2\gamma}\left(\sqrt{1+2\gamma\frac{k^2}{m
}}-1\right),
\end{equation}
where $\gamma\approx1/\varepsilon_g$ is the nonparabolic parameter,
with
 $\varepsilon_g$ as
 the energy gap between
the conduction and valence bands. Note that Kane energy band becomes
the parabolic case for vanishing $\gamma$. The nonparabolic factor
in the numerical calculation is set to be $\gamma=2.73\, {\rm
eV}^{-1}$ for InAs \cite{cao2002iii}. The $\delta$-form short-range
or the remote charged impurity scattering is considered in the
calculation. The scattering matrix of remote electron-impurity
scattering takes the form: $\vert u(q)\vert^2\simeq
n_ie^{-2sq}I(q)^2$ \cite{lei1985tdb}, with $I(q)$ as the form
factor. We set the electron effective mass at the band bottom
$m=0.04m_{\rm e}$ ($m_{\rm e}$ is the free electron
 mass), remote impurities
in InSb barrier are located at a distance of $s=10$ nm from the
interface of the heterojuction \cite{lin2006she,cao2002iii}.

\begin{figure}[t]
    \begin{center}
        \includegraphics[width=0.39\textwidth]{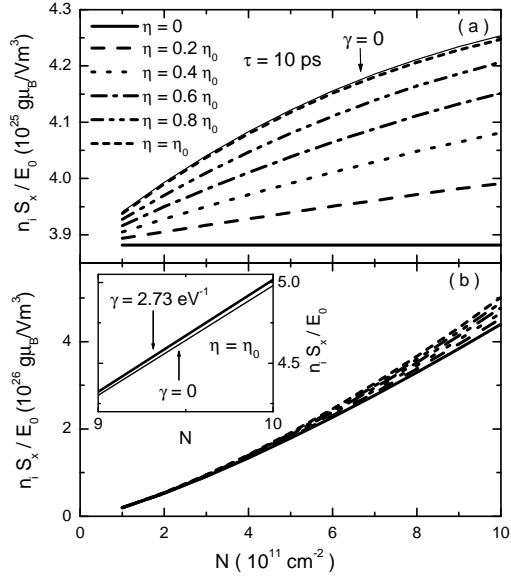}
    \end{center}
    \caption{$n_iS_x/E_0$
     as functions of electron density for (a)
      short- or (b) long-range impurity scattering for $\gamma=2.73\, {\rm eV}^{-1}$.
    Here $\eta_0=1.0\times10^{-28} {\rm eVm^3}$, ${g}$ is the effective g-factor, and $\mu_B$ is
    the Bohr magneton. The thin solid line in (a) is obtained for $\gamma=0$ and $\eta=\eta_0$.
    The thin and thick solid lines in inset of (b) is calculated when $\eta=\eta_0$ for $\gamma=0$
    and $\gamma=2.73\, {\rm eV}^{-1}$, respectively. The unit of electron density $N$ is $10^{11}\,{\rm cm^{-2}}$,
    and the unit of $n_iS_x/E_0$ is $10^{26}\,{\rm g\mu_B/Vm^3}$.}
    \label{fig:N}
\end{figure}

\subsubsection{$H_D^{(1)}+H_D^{(3)}$}
First we consider the system with linear and cubic Dresselhaus SOC.
In this case, the spin orbit field $\bm b(\bm k)=(\beta k_x+\eta k_x
k_y^2,-\beta k_y-\eta k_y k_x^2)$. At the same time, the equations
about distribution function, from Eq. \eqref{eqli} to Eq.
\eqref{eq2}, can be obtained, by substituting the new form of energy
$\varepsilon_M(\bm k)$ and $\chi_{\bm k}$ with $\varepsilon_M(\bm
k)=\sqrt{b_x(\bm k)^2+b_y(\bm k)^2}$, $\chi_{\bm
k}=-\tan^{-1}\frac{b_x(\bm k)}{b_y(\bm k)}$.
It is noted that the energy spectrum becomes anisotropic completely.

\begin{figure}[t]
    \begin{center}
        \includegraphics[width=0.39\textwidth]{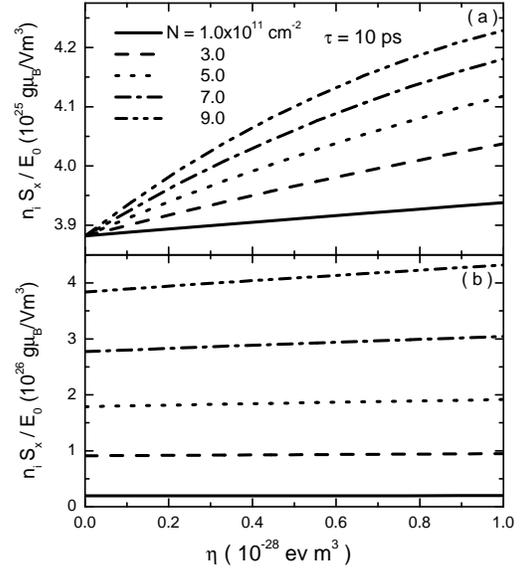}
    \end{center}
    \caption{Dependencies of $n_iS_x/E_0$ on
    cubic Dresselhaus SOC parameter for (a) short- or (b) long-range
impurity scattering.}
    \label{fig:eta}
\end{figure}

  From Eqs. \eqref{eq1} and \eqref{sxm}, we find that the $x$
  component of CISP is inverse proportional to the impurity density $n_i$.
  Therefore, $n_iS_x/E_0$ is plotted
  in these figures. In the calculation, we
  set linear Dresselhaus SOC coefficient $\beta=1.0\times10^{-11} {\rm eV
 m}$. The short-range impurity
scattering is considered with relaxation time $\tau=10\,\rm ps$.
When the cubic Dresselhaus SOC is considered, it is found that CISP
is still along the $x$ direction.

In Fig. \ref{fig:N}, the $x$ component of CISP
 is plotted as function of electron density. For short-range disorder, the density dependence of CISP can be observed
 when
 $\eta\neq0$. Note that the resultant $S_x$ is proportional to relaxation time $\tau$.
 With the increment of the sheet density, CISP
 increases monotonously, and almost saturates at high density for large $\eta$.
 From Fig. \ref{fig:N}(b), for long-range electron-impurity scattering
  it is evident that, unlike the case for short-range disorder, here CISP
 always increases with ascending the density even for the system without cubic SOC.
 In the parameter regime, $N<10^{12} {\rm cm^{-2}}$, long-range
  disorders have strong effect on CISP, where CISP increases
  rapidly with the rise of density. It can be seen
 that the role of cubic term of Dresselhaus SOC on CISP becomes
 important at high sheet density for both short- and long-range collision. CISP for parabolic energy band is
 also plotted in this figure with a thin line. We find that the weak effect of
nonparabolicity on CISP appears at high density.

 CISP is shown as a function of the cubic Dresselhaus SOC
 parameter $\eta$ in Fig. \ref{fig:eta}. For momentum independent potential, CISP
 begins with the value ${em\beta\tau E_0}/{2\pi}$, independent of the density, and increases with ascending $\eta$.
 In Fig. \ref{fig:eta}(b), the calculated CISP for long-range collision is almost linear proportional to cubic Dresselhaus constant
 $\eta$.

\subsubsection{$H_D^{(1)}+H_D^{(3)}+H_{BIA}^{(3)}$}

\begin{figure}[t]
    \begin{center}
        \includegraphics[width=0.4\textwidth]{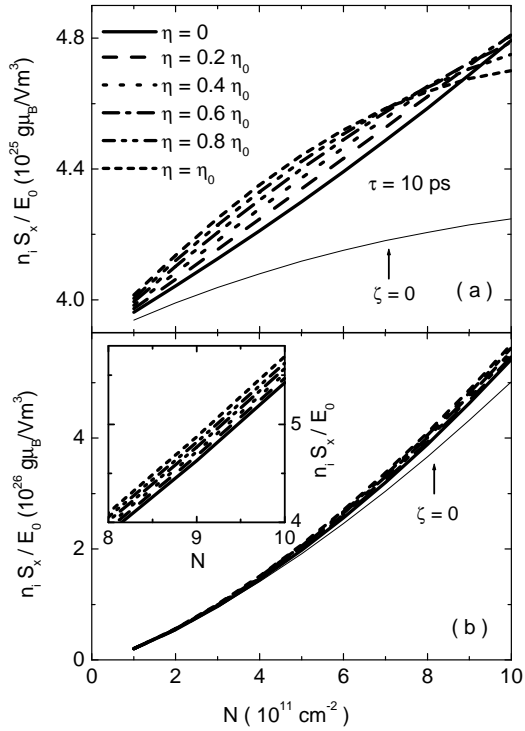}
    \end{center}
    \caption{{$n_iS_x/E_0$ is shown}
     as functions of electron density for {(a)}
      short- or {(b)} long-range impurity scattering when the additional cubic SOC term $H_{BIA}^{(3)}$ is considered.
      The thin solid lines in (a) and (b) are obtained when $\zeta=0$ and
      $\eta=\eta_0$ for short-range and long-range collisions,
      respectively. The inset in (b) shows the dependencies of
      $n_iS_x/E_0$ on $N$ at high density regime.
      The other parameters are the same as those in Figure \ref{fig:N}.
} \label{fig:Nzeta}
\end{figure}


In this subsection, the additional high-order contribution
$H_{BIA}^{(3)}$ due to bulk inversion
   asymmetry in Ref. \cite{Cartoixa2006} is
also considered. Now the spin orbit field becomes $\bm b(\bm
k)=(\beta k_x+\eta k_x k_y^2-\zeta k_x^3,-\beta k_y-\eta k_y
k_x^2+\zeta k_y^3)$, and the corresponding $\varepsilon_M{(\bm k)}$,
$\chi_{\bm k}$ can be obtained analogously. We take the well width
$a=5\,{\rm nm}$.
 The calculated $n_iS_x/E_0$ as functions of electron density $N$
is shown in Fig. \ref{fig:Nzeta}.

When the additional high-order term $H_{BIA}^{(3)}$ is included, the
magnitude of CISP rises for both short- and long-range disorders.
For short-range scattering, $n_iS_x/E_0$ always increases with
ascending the density. However, at high density, the magnitude of
CISP for large $\eta$ may be less than the one for small $\eta$.
This is due to the interplay between two cubic terms. It has been
seen that CISP saturates at high density when only the Dresselhaus
cubic term $H_D^{(3)}$ is included. However, one can see, from the
thick solid line in Fig. \ref{fig:Nzeta}(a), this behavior will not
occur when we only include the term $H^{(3)}_{BIA}$. For large
$\eta$, the effect on CISP of cubic Dresselhaus
 term $H_D^{(3)}$ exceeds the one of this additional cubic term $H_{BIA}^{(3)}$. CISP saturates again at
high density, hence its magnitude becomes less than the one with
small $\eta$. However, such phenomenon can not be observed for the
case of long-range collision.

\section{conclusion}
In summary, the CISP for 2DEG with a general SOC is investigated.
For isotropic energy band, we find that the in-plane CISP becomes
density-dependent for non-linear SOC, and increases with enhancing
the sheet density. We have numerically studied the linear and cubic
SOC contributions to CISP, considering both the short- and
long-range disorders. For short-range collision, we have
demonstrated the dependencies of CISP on density when high-order
SOCs are included. When impurity scattering becomes long ranged,
however, CISP increases rapidly with raising the density even for
the system without cubic SOC. Our investigation indicates that the
remote disorder has a strong influence on spin polarization, and the
magnitude of CISP strongly relies on the scattering matrix. The
contributions of different cubic SOCs to CISP can be comparable.

\begin{acknowledgement}
We are very thankful to N. S. Averkiev for useful information. HTC
gratefully acknowledges support from the Special Foundation of
 Theoretical Physics of NSF in China (grant 10747159).
\end{acknowledgement}

\providecommand{\WileyBibTextsc}{}
\let\textsc\WileyBibTextsc
\providecommand{\othercit}{} \providecommand{\jr}[1]{#1}
\providecommand{\etal}{~et~al.}

\end{document}